Preference for redistribution and institutional trust: Comparison before and after COVID-19

Eiji Yamamura,[a,*], and Fumio Ohtake[b]

[a]Department of Economics, Seinan Gakuin University, Japan

Email: yamaei@seinan-gu.ac.jp

[b]Center for Infectious Disease Education and Research, Osaka University, Japan,

ohtake@cider.osaka-u.ac.jp

*Corresponding author

Department of Economics, Seinan Gakuin University

6-2-92 Sawaraku Nishijin

Fukuoka 814-8511

Japan

Tel.: 81-92-823-4543

Fax.: 81-92-823-2506

Email: yamaei@seinan-gu.ac.jp

Abstract

Using an individual-level panel dataset from Japan covering the period 2016–2024, we examined how the COVID-19 pandemic, as an unanticipated public crisis, affected preferences for income redistribution. Furthermore, we investigated how the association between redistribution preferences and trust in government changed before and after COVID-19. The major findings are as follows: (1) individuals in the high-income group are less likely to prefer redistribution after COVID 19 than before it; (2) the degree of decline in redistribution preference is lower when trust in government is higher; and (3) generalised trust and reciprocity did not influence the decline in preference.

# 1. Introduction

Unexpected disasters influencing income inequality have increasingly attracted researchers' attention. However, views on the effects of disastrous events on income inequality are conflicting[1]. In historical studies of infectious diseases, the Spanish flu of 1918 was associated with an increase in inequality (Galletta and Giommoni, 2022), whereas the fourteenth-century Black Death led to a reduction in inequality (Alfani, 2022). Similar opposing effects have also been observed in the present day. Empirical findings on the impact of COVID-19 on inequality are mixed: some studies report an increase in inequality (Angelov and Waldenström, 2023; Cortes and Forsythe, 2023), while others find a reduction (Clark et al., 2021). However, contradictory outcomes may not, in fact, be inconsistent.

Disaster victims typically experience economic hardships. Income inequality increases if the impact of disasters is severe for low-income households than for high-income households (Sologon et al., 2022). Low-wage workers are more likely to face economic strain than high-wage workers (Dalton et al., 2021). This disparity may be partly attributable to the fact that excess mortality rates are higher among lower-income groups(Decoster et al., 2021). However, governments have implemented income distribution policies to provide relief to disaster victims, thus playing a cushioning role in response to COVID-19 (Almeida, 2021; Brewer and Tasseva, 2021; Sologon et al., 2022). If subsidies for disaster relief are sufficiently large, income inequality may even decline relative to pre-disaster levels.

---

[1] As a consequence of natural disasters, income inequality increased (Yamamura, 2015), whereas the inequality decreased (Vignoboul, 2025).

In examining redistribution policies, we should consider the possibility that moral hazards may increase. In the empirical work, Empirical studies suggest that preferences for redistribution are influenced by perceptions of public morality(Algan et al., 2016) and belief about fairness (Fong 2001; Roth and Wohlfart, 2018; Sabatini et al., 2020) [2]. Trust and confidence in government lead individuals to increase redistribution preferences (e.g., Kuziemko et al., 2015; Yamamura, 2014a)[3]. However, income redistribution policy might cause a problem if public morality is weak. On the one hand, the disasters may incentivise bureaucratic corruption (Yamamura, 2014b). On the other hand, income redistribution policies to mitigate damage from disasters may motivate people to engage in fraudulent behaviour. Consequently, institutional trust declines[4]. If trust erodes, long-term public support for government-led redistribution may decline, potentially resulting in a vicious cycle. It is critical not to fall into this vicious circle. In countries characterised by high trust and low corruption, individuals support high taxation (Algan et al., 2016, 862).

Following COVID-19 in Japan, various types of subsidies have been used to assist people experiencing economic hardships. However, the emergence of fraudulent claims has become a major social issue[5]. The primary purpose of this study was to examine whether COVID-19 reduced redistribution preferences, taking into account institutional

---

[2] Many studies have explored the determinants of redistribution preference (e.g. Alesina and La Ferrara, 2005; Bellani and Scervini, 2015, 2020; Corneo and Grüner, 2002; Luttmer and Singhal, 2011; Ravallion and Lokshin, 2000; Yamamura, 2012, 2025).
[3] Trust in government was critical to coping with the COVID-19 pandemic (Lazarus et al., 2020).
[4] During the swine flu epidemic (H1N1), institutional trust declined at the peak of the crisis (van der Weerd et al., 2011). In Sweden, short-term analyses indicate that COVID-19 increased institutional trust (Esaiasson et al., 2020).
[5] Experimental study during COVID-19 found that trust level became below the baseline level (Shachat et al., 2021). Meanwhile, natural disasters increased social capital such as generalised trust and then subjective well-being (Yamamura et al., 2014, Helliwell et al., 2023).

trust, based on data from Japan.

This study is closely aligned with recent research examining the relationship between institutional trust and redistribution preference influenced by disastrous shocks, such as the Russo-Ukrainian war (Zakharov and Chapkovski, 2025) and COVID-19 pandemic. Daniele et al. (2020) conducted a survey experiment in European countries to investigate the influence of the perceived consequences of COVID-19 and found that it reduced institutional and generalised trust and support for redistribution policies. Using a panel dataset from Germany, Bellani et al. (2023) found that the severity of the COVID-19 infection rate negatively influenced redistribution preferences, mediated by a drop in institutional trust. However, these analyses primarily focused on short-term effects.

Recent evidence suggests that the impacts of COVID-19 have been persistent, particularly among vulnerable populations (Neidhöfer et al., 2021). From a long-term viewpoint, some studies have explored the influence of major epidemics on income distribution (Brzezinski, 2021; Furceri et al., 2022). Saka et al. (2022) found that COVID-19 induced a long-term erosion of public trust in government. Nevertheless, these studies did not cover the post-COVID-19 pandemic period. It is unclear whether the influence of COVID-19 on redistribution preference and trust in the government persisted after thepandemic had largely subsided.

An independently constructed individual-level panel dataset for the period 2016–2024 enabled us to compare redistribution preferences before and after COVID 19by tracking the same individuals over time. This is an advantage of this study. We found that individuals in the high-income group became less likely to prefer redistribution after COVID-19 than before it. However, institutional trust played a critical role in maintaining a high-income redistribution preference. The central contribution of this study lies in demonstrating that the COVID-19 pandemic exerted a lasting influence on redistribution preferences, operating through a sustained reduction in institutional trust, even after the acute phase of the crisis had passed.

The remainder of this paper is organised as follows: Section 2 describes the real situation in Japan. Section 3 describes the data used in this study. Section 4 proposes testable hypotheses and describes the empirical method. Section 5 presents the estimation results and their interpretations. Section 6 provides concluding remarks.

## 2. Government Response and Public Trust During the COVID-19 Pandemic in Japan

Japan reported its first cases of COVID-19 in March 2020. In response to the crisis, state of emergency was declared, and the public was asked to refrain from going out and avoid face-to-face exchanges and communication. Restaurants were asked to shorten their business hours. In addition, various entertainment events were asked to be cancelled[6]. The Tokyo Olympics, scheduled for the summer of 2020, were postponed, resulting in significant losses to the tourism and restaurant sectors. In Japan's 2020 fiscal year, the total expenditure on supplementary budgets amounted to approximately 77 trillion yen in additional spending, with the corresponding additional bond issuances totalling approximately 80 trillion yen.

The Japanese government implemented various measures to mitigate the impact of COVID-19 on the economy. The government expanded its regular tax deferral system.

---

[6] However, in Japan, people were not fined even if they went out during the state of emergency. In Italy, during the lockdown, higher fines for non-compliance with restrictions were associated with lower levels of trust in the government (Fazio et al., 2022).

For example, individuals whose income decreased by more than 20 % due to the COVID-19 crisis or who temporarily faced difficulties in paying taxes were granted tax deferrals and exemptions from late payment penalties (Nakata, 2023, 236).

In particular, various types of subsidies were newly implemented to support those experiencing financial hardship due to COVID-19. For instance, the Employment Adjustment Subsidy is a system that subsidizes the costs incurred by business owners forced to downsize their business activities for economic reasons in order to maintain employment[7]. Since the COVID-19 outbreak, the prefectural government has published the names of companies that have fraudulently received employment adjustment subsidies. According to Tokyo Shoko Research (2025), the total number of cases of the fraudulent receipt of employment adjustment subsidies has reached 1,620 since April 2020. The total amount of fraudulent receipts reached 53.04 billion yen.

Sustainability subsidies support the continuation of businesses particularly affected by the spread of infectious diseases and businesses particularly affected by the pandemic. So, the subsidies could be used flexibly across various business activities. The Ministry of Economy, Trade, and Industry has published information on its official website about those who have fraudulently received Business Continuity Subsidies. In May 2025, the Small and Medium Enterprise Agency announced 2,439 cases of fraudulent receipt of benefits, totalling over 2.4 billion yen since 2020[8].

Fig. 1 reveals that online search activity for 'fraudulent receipt' on the internet spiked between March 2020 to September 2020. This was considered to reflect a real situation.

---

[7] Official site of the Ministry of Health Labour and Welfare.
https://www.mhlw.go.jp/stf/seisakunitsuite/bunya/koyou_roudou/koyou/kyufukin/pageL07_20200515.html.
(Accessed on 4 July 4 2025).

[8] https://www.meti.go.jp/covid-19/jizokuka_fusei_nintei.html. (Accessed on 4 July 4 2025).

In April 2020, as part of the first economic response to the onset of COVID-19, the government introduced emergency measures that included a grant of 300,000 yen per household for those whose income had fallen below a certain threshold. However, the identification and verification process for eligible households requires a tremendous amount of time and money. Criticism of this policy escalated because it was not an effective way to deal with emergencies. Inevitably, this initial proposal was withdrawn. Instead, the government adopted a policy in which all households could receive a special fixed-benefit payment of 100,000 yen per person. Subsequently, the government implemented various types of subsidies.

Fig. 1 shows that the Japanese paid a great attention to fraudulent receipts in response to implementing various subsidies. With increased public awareness of the scale of fraud, trust in the Japanese government was perceived to have declined.

3. Data and method

3.1. Data

To construct the panel dataset, internet surveys were conducted by sending questionnaires to the same participants. For this study, we divided the study period into before and after COVID-19 outbreak. Prior to COVID-19, we gathered data in 2016 (12–19 July), 2017 (11–19 July), and 2018 (3–11 October). After the outbreak, data were collected in 2021 (2–8 December), 2023 (2–9 March), and 2024 (1–9 February). In this way, we independently constructed a panel dataset. Nonetheless, some participants did not respond to the surveys and inevitably dropped out. To maintain the sample size, new participants were recruited for subsequent surveys. We primarily used balanced panel data, comprising the same individuals who participated in all surveys. An unbalanced panel was also used for robustness check.

Prior to the initial survey in 2016, the Nikkei Research Company (NRC) was selected to conduct the survey, as NRC is known for undertaking various types of academic surveys. We commissioned the NRC to conduct surveys until a sufficient number of samples were collected[9].

After the subsequent survey, the NRC collected data by tracking the same respondents to construct panel dataset. In the second survey, conducted in 2017, we gathered 9,130 observations, including newly recruited respondents. Then, respondents from 2017 were matched with those from 2016. Specifically, 7,107 respondents participated in both surveys, while 2,023 were new participants. Following this procedure, we gathered observations in 2018, 2021, 2023, and 2024. In the estimations, participants who did not respond to questions regarding the control variables were excluded from the sample.

Using balanced panel data, in which respondents appeared in all surveys, the number of observations was reduced to approximately 10,000. However, some participants were excluded from the specification where 'Reciprocity' was used as a control variable, reducing observations to approximately 8,600. This was because 'Reciprocity' question was omitted from the 2024 survey due to strict budget constraints The unbalanced panel data included approximately 27,000 observations when 'Reciprocity' was not included, and 24,000 when it was.

The questionnaire included questions about the demographic characteristics of the respondents, such as gender and birth year. For economic and social status, we included variables such as annual household income, job status, and educational background.

---

[9] In the dataset, respondents were divided into five cohorts (20-29, 30-39, 40-49, 50-59, and over 60) and according to sex. This resulted in 10 demographic groups: five male cohorts and five females cohorts. Further, these were divided into six regions (Hokkaido-Tohoku, Kanto, Chubu, Kinki, Shikoku-Chugoku, Kyushu-Okinawa). Therefore, respondents were categorised into 60 groups. We planned to gather observations almost equivalent to the population share in each category of the Japanese Census. Hence, the sample gathered in this study is considered to be representative of the population structure of Japan.

Questions related to subjective values were also included, as these served key variables. The dependent variable was redistribution preference. Drawing on established survey items from the literature (e.g., Alesina and La Ferrara, 2005; Bellani and Scervini, 2024; Corneo and Grüner, 2002; Ravallion and Lokshin, 2000), we asked respondents, 'Please indicate your opinion on the following statements from 5 choices', 'Agree' (assigned a value of 1) to 'Agree' (assigned a value of 5).

.

*"The government should reduce the gap between the rich and poor"*

The key independent variables were trust in government, generalised trust, and views about reciprocity. The questions were as follows:

*'How much trust do you have in the government?'*

*'Do you generally trust people?'*

*'If someone has helped you in the past, do you feel you should help them in return?'*

For the trust-related questions, respondents selected from five options ranging from 'I don't trust at all' (assigned a value of 1) to 'I trust very much' (assigned a value of 5). Regarding reciprocity, there were five choices ranging from 'I do not agree at all' (assigned a value of 1) to 'I strongly agree' (assigned a value of 5).

The primary variable of interest in this study is trust in the government. However, various types of pro-social values may be correlated with redistribution preferences. For instance, generalised trust fosters a civic society grounded in welfare states (Algan et al., 2016; Daniele and Geys, 2015). Reciprocity is thought to enhance mutual support during hardship and is positively correlated with preference for redistribution (Fong et al., 2006). Therefore, we also included proxy variables for generalised trust and reciprocity.

## 3.2. Analysis using raw data

Table 1 presents the definitions of the variables used in the regression estimations and their mean values before and after COVID 19, based on a balanced panel. *Redistribution* reduced from 3.61 to 3.60 after the COVID-19 outbreak. The degree of reduction was marginal. This is reasonable because income distribution policies benefit low-income earners but impose a burden on high-income earners. Inevitably, the opposite effects cancel each other out, resulting in a slight overall change. After COVID-19, the proportion of respondents who reported high trust in government (choosing 4 or 5) slightly decreased from 0.18 and 0.03 to 0.17 and 0.02, respectively, while that of those who do not trust the government (Choosing 2) increased from 0.21 to 0.25. Overall, trust in government declined. A similar trend was observed in the composition of generalised trust.

Fig. 2 shows the household income level according to the redistribution preference that respondents chose before and after COVID- 19. The mean income level of those who prefer redistribution is slightly higher than that of those who do not. High-income earners may prefer redistribution to avoid conflicts between high- and low-income earners (Yamamura, 2016a). However, after COVID-19, the mean income level increased among those who did not prefer redistribution. COVID-19 is thought to reduce high-income earners' support for redistribution.

Figs. 3(a)-(c) illustrate the means of trust in the government, generalised trust, and reciprocity by redistribution preferences. In Fig. 3(a), prior to COVID-19, the mean trust

in government is the lowest for those who chose 1 and do not prefer redistribution. The value almost increases as the group becomes more positive about redistribution. However, after COVID-19, people who did not prefer redistribution showed higher trust in the government than those who preferred redistribution. Hence, the positive relationship between redistribution preference and trust in the government disappeared after the respondents experienced COVID-19. In contrast, in Fig. 3(b), the positive relationship between generalised trust and redistribution preference persisted in both periods, although the mean values slightly declined for those who preferred redistribution. Similarly, Fig. 3(c) shows a positive relationship between reciprocity and redistribution preference. Comparing Figs. 3(a)-(c), COVID-19 had a greater effect on the relationship between trust in the government and redistribution preference than on other relations.

## 4. Hypotheses and method

To reduce income inequality, the government allocates wealth from high to low-income earners, through various measures. Imposing higher taxes on high-income earners improves the economic condition of low-income earners. After disastrous events, government implemented various measures to mitigate income inequality. This may increase the burden on high-income earners, leading to a decrease in their net benefit from the redistribution policy. Accordingly, *Hypothesis 1* was proposed:

*Hypothesis 1:*

*Rich people are less inclined to prefer income redistribution after COVID 19 than before COVID 19.*

However, we can add the assumption that redistribution preference depends on beliefs about distributive justice rather than self-interest (Fong, 2001). High- income individuals may support redistribution more than low- income individuals if they believe low income is determined by circumstances beyond individual control. The redistributive mechanism could work when high-income earners expect to be surrounded by honest individuals (Algan et al., 2016).

The occurrence of moral hazard problems such as fraudulent receipts undermines fairness. The probability of the fraudulent receipt of subsidies seems to depend considerably on the quality of the government. This influences taxpayer perceptions (Fehr and Schmidt, 1999; Alesina and Angeletos, 2005). Consequently, people's distrust in the government reduces their willingness to pay taxes (Oh and Hong, 2012). In other words, institutional trust is important for high-income earners to support income redistribution at their own expense.

The effects of social conditions, such as beliefs and trust, differ according to income level. Redistribution preference is influenced by social capital (Yamamura, 2012) and trust in the government (Yamamura, 2014a), which is observed only for people in the high-income group but not for those in the low-income one.

As illustrated in Fig. 4, before the outbreak of COVID-19, support for redistribution policy was positively correlated with income level, which cannot be explained by an individual's self-interest, because an individual's net benefit from redistribution policy increases as income decreases. Surprisingly, after COVID-19, we observed that redistribution preference clearly declined as income levels increased, which can be explained by trust rather than self-interest. Furthermore, at income levels above the median (500), redistribution preference tended to be lower after COVID-19 than before it.

As shown in Figs. 5(a) and (b), high-income earners were more likely to trust the government and others than were low-income individuals before COVID-19 than after it. Table 1 shows that trust decreased after the COVID-19 outbreak, which is consistent with previous studies (Bellani et al. 2023; Daniele et al., 2020; Hensel et al., 2022; Saka et al., 2022) but not in line with Esaiasson et al. (2020). In addition, the reduction is observed for the group over the median income (500 thousand yen) but is hardly observed below it. That is, the gap in trust between poor and rich people reduced after COVID-19. In contrast, in Fig. 5(c), high-income earners become more likely to be reciprocal after COVID-19. One reason might be that in the question of reciprocity, people were asked about their response to the assumption that someone had helped them in the past. Meanwhile, for the question about trust, no assumption was made. Therefore, fraudulent behaviour in subsidies hardly influenced reciprocity.

High institutional trust was considered to neutralise high-earner's negative views about redistribution. However, a divergence of institutional trust among high-income earners occurred during the COVID-19 period. They may possibly find reasons to blame their government for inefficient responses to natural disasters (Healy and Malhotra, 2009). Meanwhile, their redistribution preferences persist if they continue to trust the government. Thus, even at the same income level, the redistribution preferences of higher earners vary according to their institutional trust. Hence, we proposed *Hypothesis 2*:

*Hypothesis 2:*

*The degree of decline in rich people's redistribution preference is smaller if their trust in government is stronger.*

## 4.1. Method

The baseline model assessed *Hypothesis 1* as follows:

$$Redistribution_{it} = \alpha_0 + \alpha_1 \, COVID_t \times High\ Income_{it} + \alpha_2 \, COVID_t + \alpha_3 \, High\ Income_{it} + X'_{it} B + u_i + e_{it} \quad (1)$$

*Redistribution*$_i$ is the dependent variable. α denotes the coefficients of the variables. *i* and *t* represent individuals and time points, respectively. X is the vector of the control variables and B is the vector of their coefficients. X represents age, marital status, job status, residential prefecture, and household income. Alternative specifications include dummies for government trust, generalised trust, and reciprocity as control variables. The Fixed-Effects model controls for an individual's time-invariant characteristics such as gender and educational background, represented by $u_i$. $e_{it}$ is an error term.

In Section 3, Figs. 4 and 5 show that median income is the threshold for the reversal of income distribution preferences before and after COVID-19. The effect of income should be captured using a binary rather than a linear variable. Hence, we used *High Income,* which is a dummy variable for those whose household income is higher than the median household income.

The key independent variable was $COVID_t \times High\ Income$, which is the interaction term between *COVID* and *High Income.* From *Hypothesis 1*, the sign of the coefficient of $COVID_t \times High\ Income$ was predicted to be negative. In the model, *High Income* shows how the redistribution preferences of high-income earners differed from those of low-income ones before *COVID-19*. Furthermore, apart from the main results, we reported a linear combination $COVID \times High\ Income + High\ Income$, indicating a difference in the redistribution preference of high-income earners from low-income ones after *COVID-19*.

To test *the Hypothesis 2,* we added triple interaction terms to the baseline model as follows:

$$Redistribution_{it} = \beta_1 COVID_t \times High\ Income_{it} \times Government\ Trust_{it}$$
$$+ \beta_2 COVID_t \times High\ Income_{it} \times General\ Trust_{it} + \beta_3 COVID_t \times High\ Income_{it} \times$$
$$Reciprocity_{it} + Z'_{it}C + u_i + e_{it} \quad (2)$$

As shown in Fig. 3(a), the relationship between *Government Trust* and *Redistribution Preference* is not linear. Therefore, concerning subjective values, Government Trust, General Trust, and Reciprocity, we used four dummy variables where the group of those who chose '1' is the reference group. For *Government Trust, COVID×High Income× Government Trust_2, COVID×High Income×Government Trust_3, COVID×High Income×Government Trust_4,* and *COVID×High Income×Government Trust_5* were added. From *Hypothesis 2,* the expected signs of their coefficients are positive and their absolute values are larger if they interact with higher trust dummies. Furthermore, as a linear combination, *Covid ×High Income+ Covid ×High Income× Government Trust_5*.

Further, we added three interaction terms: *COVID × High Income, COVID × Government Trust*, and *High Income×Government Trust*. Taking the example of *Trust_2*, we added *COVID×Government Trust_2* and *High Income×Government Trust_2*. In addition, the interaction terms *Trust_3*, *Trust_4*, and *Trust_5* were included. However, the results of these variables are not reported in Tables 4 and 5 because they were not directly used to test *Hypothesis 2*.

We considered whether other pro-social factors were correlated with changes in redistribution preference after COVID-19. For this purpose, we also incorporated General Trust and Reciprocity to consider other channels that influence redistribution. In other words, we added various interaction terms for *General Trust* and *Reciprocity*.

## 5. Estimation results and discussion

### 5.1. Estimation results

First, we examine the visual relationship between household income and redistribution preference. In Fig. 6, we present scatter plots of the mean values of household income and redistribution preference for 47 residential prefectures, before and after COVID-19. Fig. 6 shows a positive association before COVID-19 and a negative association after COVID-19. This finding is congruent with *Hypothesis 1*. However, in Fig. 6, the differences in the characteristics of the prefectures and other factors are not controlled. For a closer examination, we refer to Tables 2 and 3.

Table 2 shows results using balanced panel data, while Table 3 presents those using unbalanced panel data. In Table 2, consistent with *Hypothesis 1*, the coefficient of *COVID ×High Income* shows the predicted negative signs and statistical significance at the 1 % level in all columns. The absolute value of the coefficient of *COVID ×High Income* is approximately 0.3. In our interpretation, in comparison with the lower-income group, respondents in the high-income group are less likely to prefer redistribution by 0.30 point on a five-point scale after COVID-19 compared to before. A significant positive sign of *High Income* indicates that people whose income was above the median value are more likely to prefer redistribution than those with lower income before COVID-19. The absolute values of the coefficient are 0.14–0.17. As for linear combination, the sign of *COVID×High Income + High Income* is negative and statistically significant at the 1 % level in all columns. This implies that people whose income is above the median value are less likely to prefer redistribution than those with lower income after COVID-19. Further, its absolute value is approximately 0.13–0.16. Hence, the difference is almost the same, but the gap is reversed between the high- and low-income groups. This is consistent with Fig. 6.

Concerning Government Trust, all results show positive values and are statistically significant when the lowest trust group is the reference. Further, in Column (5), all dummies for subjective values are included in the model; the absolute values of the dummies for Government Trust are about 0.08 for the group of those who chose 2 or 3 to respond to the question about government trust. The values increase remarkably to 0.17 and 0.36 for those who chose 4 and 5, respectively. Individuals who prefer redistribution tend to have higher trust in the government.

For General Trust, significant negative signs are observed for those who chose 2, 3 and 4, but not for those who chose 5. Thus, the results were inconsistent. Individuals who generally trusted others did not prefer redistribution. This indicates that the correlation between government trust and redistribution preference is opposite to that between generalised trust and redistribution preference. We interpret this as follows. Unlike trust in the government, generalised trust is the foundation that enables a market economy to function (Zak and Knack, 2000). In response to emergencies, people who generally trust others place more importance on market mechanisms than on government distribution policies. As for Reciprocity, significant positive signs are observed. Furthermore, the absolute values of the coefficient increased as individuals became more likely to be reciprocal. Reciprocal people consider mutual assistance and support. Reducing income inequality is important to cultivate a mutually beneficial society. Thus, reciprocal individuals prefer the redistribution.

The results in Table 3 are almost identical to those in Table 2. The results of the baseline model are robust to alternative samples; thus, *Hypothesis 1* was strongly supported.

The results of the triple interaction terms are presented in Table 4 using a balanced sample and in Table 5 using an unbalanced panel. In all columns of Table 4, the interaction terms with dummies for *Government Trust* show the predicted positive sign. Further, statistical significance is observed for *COVID × High Income* interacting with *Government Trust_4* and *Government Trust_5,* but not with *Government Trust_2* and *Government Trust_3*. This implies that high-income individuals who trust the government are more likely to prefer redistribution than those who do not trust the government after COVID-19. This finding supports *Hypothesis 2*.

For closer examination, we consider whether high-income earners became less likely to prefer redistribution after COVID-19 than before it. For this purpose, we report a linear combination of *COVID×High Income* and its interaction with *Government Trust 4* or *Government Trust 5*. We observe a positive sign of a combination '*COVID×High Income + COVID × High Income × Government Trust_5'* in all results, despite not being statistically significant in Columns (3) and (4). We observe both positive and negative signs of '*COVID×High Income + COVID×High Income ×Government Trust_4'*. In our interpretation, high-income earners who trusted the government did not change their redistribution preferences after COVID-19. Hence, the reduction in trust in the government is the main reason why high-income earners became less likely to support income redistribution by the government after COVID-19.

Other interaction terms with *General Trust* and *Reciprocity* are not statistically significant in any column, but indicate a negative sign in most of the results. Therefore, the gap in redistribution preferences of high-income earners before and after COVID-19 did not vary according to generalised trust and reciprocity. That is, other pro-social variables did not influence the reduction in redistribution preference of high-income earners.

As shown in Table 5, the results for *COVID×Government Trust_4* and *COVID×Government Trust_5* are almost the same as those in Table 4. Hence, the results are robust when unbalanced panel data are used. In contrast to Table 4, significant positive signs are shown in the interaction terms with the dummies for pro-social values. However, statistical significance is observed only for the interaction terms with *General Trust_2* and *Trust_3* and not for those with *General Trust_4* and *Trust_5*. High-income earners with high generalised trust show reduced redistribution preference after COVID-19. The interaction with Reciprocity shows a statistically positive sign only in Column (3) and not in Column (4). Overall, Hypothesis 2 is strongly supported, as shown in Tables 4 and 5.

The combined results in Tables 2-5 indicate that high-income earners showed on average lower redistribution preference than low-income earners after COVID-19. However, this tendency is not observed among high-income earners who trusted the government. This is consistent with a panel survey (Bellani, 2023) and an experimental study (Daniele et al., 2020). Meanwhile, the reduction in high-income earners' redistribution preferences is not influenced by their other pro-social values, such as generalised trust and reciprocity. As shown in Fig. 6, before COVID-19, high-income individuals are slightly more likely to prefer redistribution than low-income individuals. This is consistent with the results of an experimental study conducted in Japan in 2018 (Yamamura, 2024). However, after COVID-19, individuals' redistribution preferences declined as their income level increased (Fig. 6).

## 5.2. Discussion

Natural disasters have a negative impact on society. Nevertheless, they may enhance interpersonal ties and volunteer activities (Yamamura, 2016b). The positive correlation between social trust and subjective well-being strengthens after natural disasters (Yamamura et al., 2014; Helliwell et al., 2023).

In contrast to natural disasters, the outbreak of pandemic exposed people to invisible threats rather than visible physical impacts. Furthermore, people were prohibited from engaging in face-to-face interactions and communication to mitigate the spread of COVID-19. Naturally, people became suspicious of one other. Even with the current reliance on online communication, the exchange of information is limited, making it difficult to foster interpersonal relationships and build trust. In this context, only a small number of people were fraudulently receiving benefits, which might have undermined their trust in the government.

The fourteenth-century Black Death is generally believed to have reduced economic inequality (Alfani, 2022). Meanwhile, seventeenth-century plagues reduced wealth inequality through income redistribution, but the effect was limited and extremely short-lived (Alfani, 2022). From our findings, the outcomes of historical epidemics may vary due to different levels of institutional trust and support for redistributive policies.

The findings of this study make it evident that high-income earners' institutional trust, such as trust in the government, is critical to supporting redistribution policies. However, moral hazard problems such as fraudulent receipts lead high-income earners to reduce institutional trust. That is, 'people are willing to help the poor, but they withdraw support when they perceive that the poor may cheat or fail to cooperate by not trying hard enough to be self-sufficient and morally upstanding' (Fong et al., 2006, p.1442). Previous studies have found that COVID-19 undermined trust and accordingly reduced redistribution preferences. The detrimental effect of COVID-19 differed according to the level of trust (Bellani et al., 2023). We placed greater emphasis on the difference between high- and low-income groups, as high-income earners may hold negative views toward income redistribution if they act in self-interest. Rich people might be motivated to prefer income redistribution if they evaluate redistribution policies based on a sense of public morality (Algan et al., 2016) and a belief in fairness (Fong, 2001; Roth and Wohlfart, 2018; Sabatini et al., 2020). What we found is consistent with it.

Furthermore, previous studies have focused on the severity of the damage suffered because of COVID-19 in the short term (Bellani et al. 2023; Daniele et al., 2020). The effect of COVID-19 may persist (Saka et al., 2022), possibly increasing tension and conflict between high- and low-income groups. Hence, it is critical for policymakers to prevent fraudulent behaviour when adopting income redistribution policies. A harmonious society depends on fairness, which is maintained by the members of society.

## 6. Conclusion

The contribution of this study is the identification of a different impact of COVID-19 on preference for redistribution between individuals with an income over median and others, which has persisted even after COVID-19 largely subsided. High-income earners became less likely to prefer income redistribution than low-income ones after COVID-19 even though they were more likely to prefer it before the pandemic. This was mainly because high-income earners distrusted the government after COVID-19. However, higher-income earners did not change their redistribution preferences if they continued to trust the government.

The occurrence of disastrous events increases income inequality. In the emergent situation, the government plays a great role in income redistribution. However, individuals and firms are motivated to fraudulently receive subsidies. There may be a trade-off between income redistribution policies and increasing unfair benefits. This may reduce institutional trust. Based on the findings of this study, we argue the following. In the short term, after a disastrous event, income inequality is mitigated through policies. In the long run, due to distrust in the government, the redistribution policy would not be supported, and income inequality would increase. The critical point is preventing fraudulent behaviour when a redistribution policy is implemented.

Due to data limitations, we could not identify causality between trust in the government and redistribution preference. Further, we assume that fraudulent receipts change the views of high-income earners about the government. However, it remains unclear whether obtaining information about fraudulent behaviours changes individuals' institutional trust. More appropriate data are needed to establish this link. Furthermore, this study is based on data from Japan. Future research should examine whether these findings hold in other countries. These limitations present important avenues for further investigation.

Journal of Public Economics, 167, 251‑262.

Sabatini, F., Ventura, M., Yamamura, E., Zamparelli, L. (2020). Fairness and the unselfish demand for redistribution by taxpayers and welfare recipients. Southern Economic Journal, 86 (3), 971‑988.

Saka, O., Eichengreen, B., Aksoy, C. (2022). The political scar of epidemics. Economic Journal, 134(660), 1683–1700.

Shachat, J., Walker M.J., Wei, L. (2021). How the onset of the Covid-19 pandemic impacted pro-social behaviour and individual preferences: Experimental evidence from China. Journal of Economic Behavior & Organization 190(C) 480-494.

Sologon, Denisa M., O'Donoghue, Cathal., Kyzyma, Iryna., Li, Jinjing., Linden, Jules., Wagener., Raymond. 2022. The COVID-19 resilience of a continental welfare regime - nowcasting the distributional impact of the crisis. Journal of Economic Inequality, 20(4), 777-809.

Tokyo Shoko Research (2025). 11th Survey of Companies Found to Have Improperly Received Employment Adjustment Subsidy. (11kai Koyo Chosei-kin Fusei Jukyu Chocsa (in Japanese)). TSR Research. https://www.tsr-net.co.jp/data/detail/1201190_1527.html

van der Weerd, W., Timmermans, D. R. M., Beaujean, D. J. M. A., Oudhoff, J. & van Steenbergen, J. E. (2011). Monitoring the level of government trust, risk perception and intention of the general public to adopt protective measures during the influenza A (H1N1) pandemic in the Netherlands. BMC Public Health 11(1): 575. https://doi.org/10.1186/1471-2458-11-575.

Vignoboul, Aubin, (2025). The winds of inequalities: How hurricanes affect inequalities
28

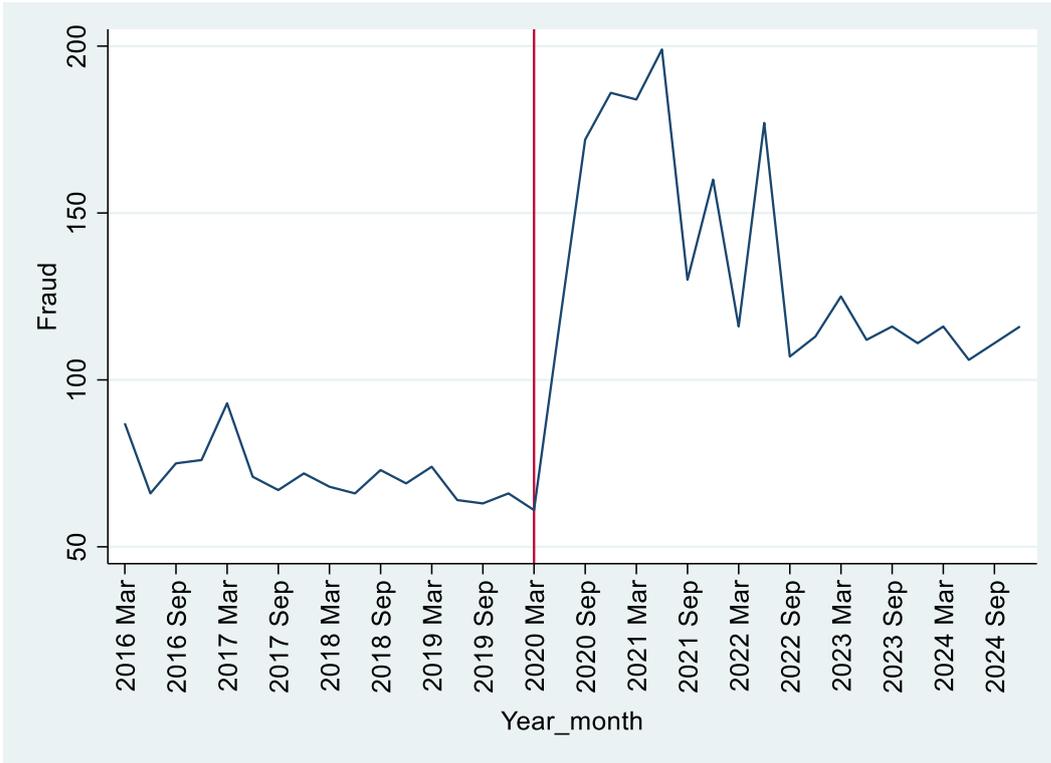

Fig. 1. Number of searches for 'Fraudulent receipt'

Note: Data were provided by Google Trends.

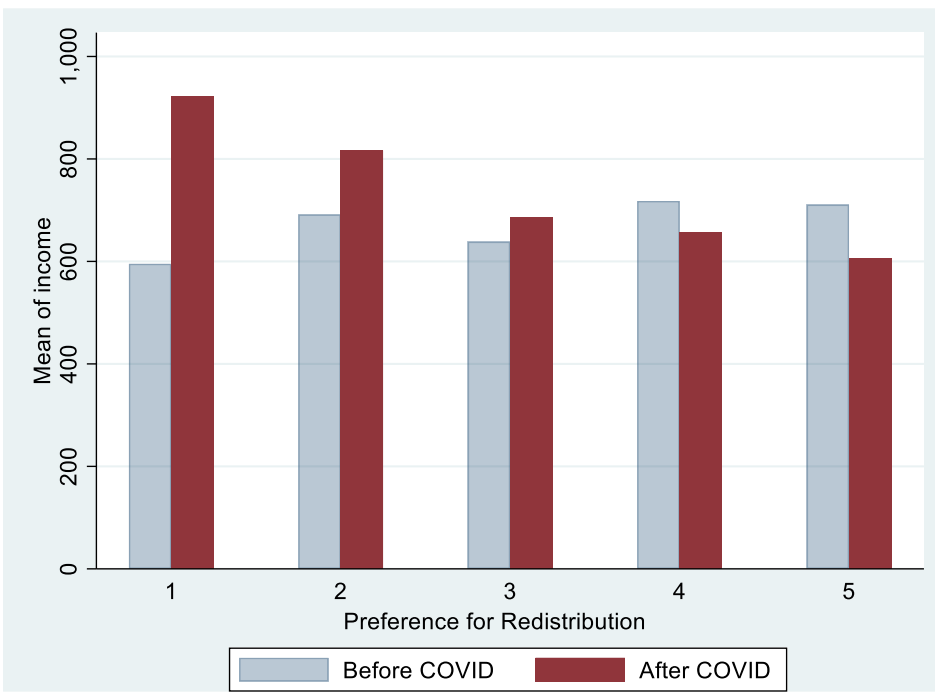

Fig. 2. Mean household income by Redistribution Preference category before and after COVID-19

Note: Balanced panel data were used.

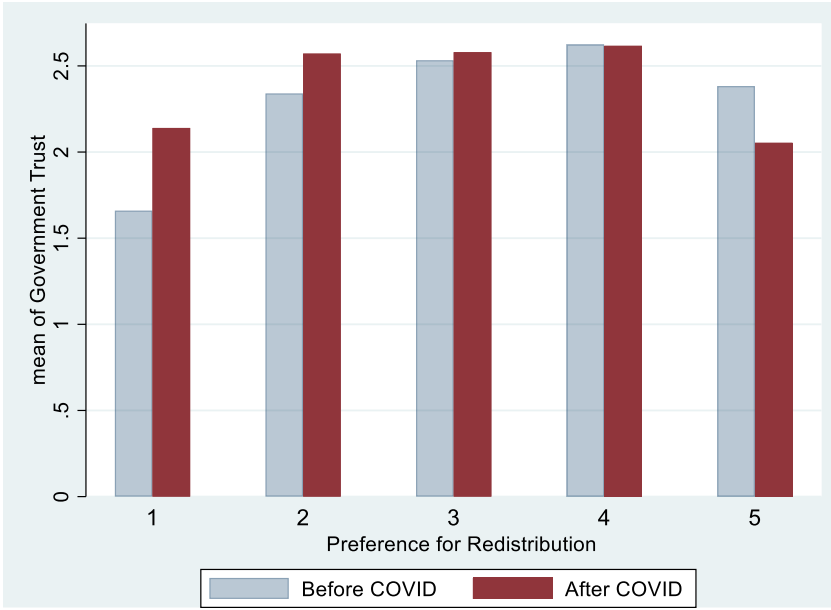

Fig. 3(a). Government trust

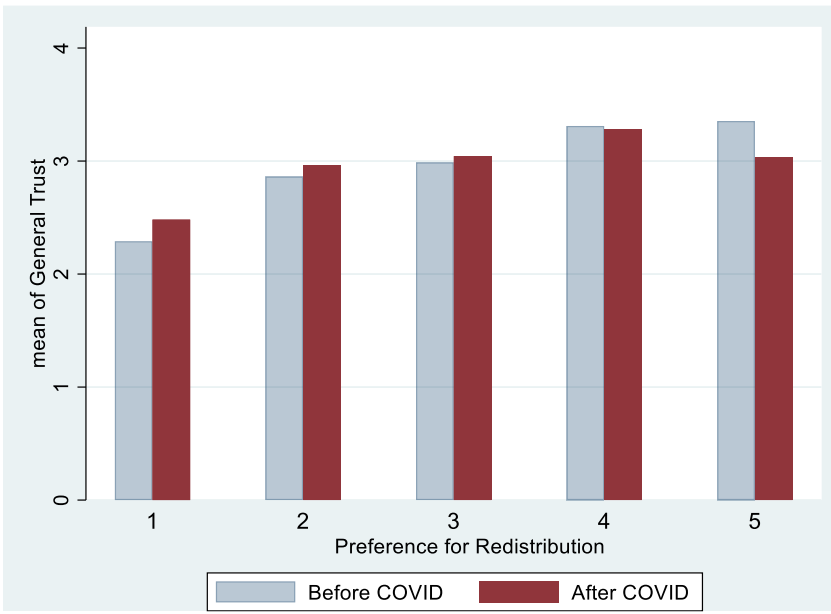

Fig. 3(b). General trust

Fig. 3. Mean subjective values by Redistribution Preference category before and after COVID-19

Note: Balanced panel data were used.

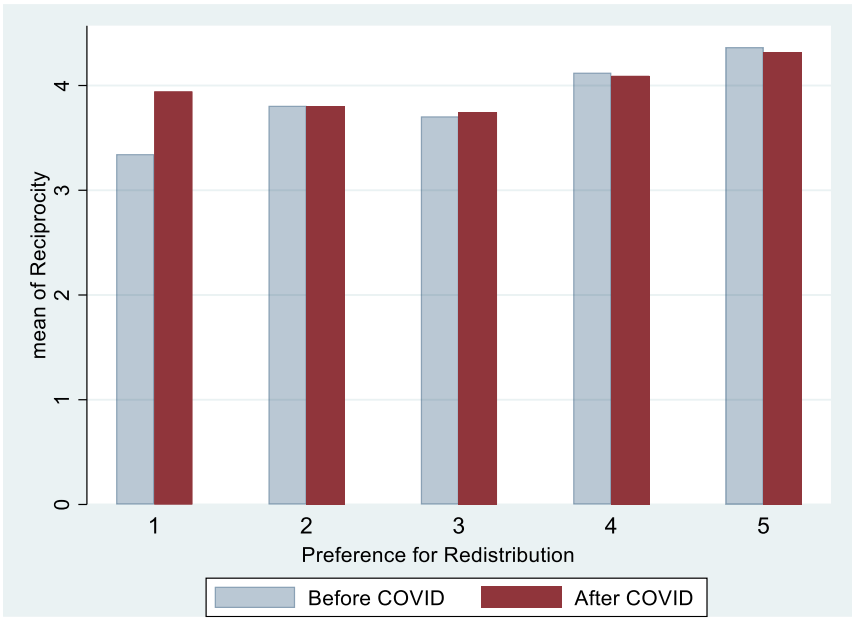

Fig. 3(c). Reciprocity.

Note: Balanced panel data were used.

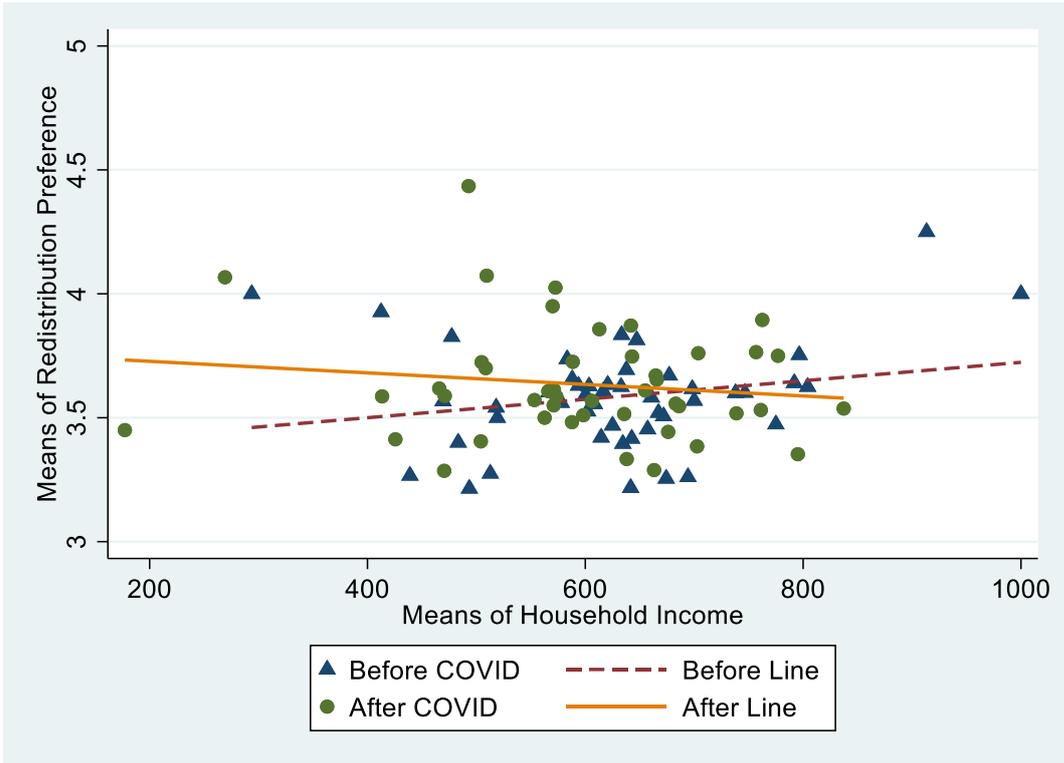

Fig. 4. Relationship between household income and Redistribution Preference before and after COVID-19 (Mean values of residential prefecture)

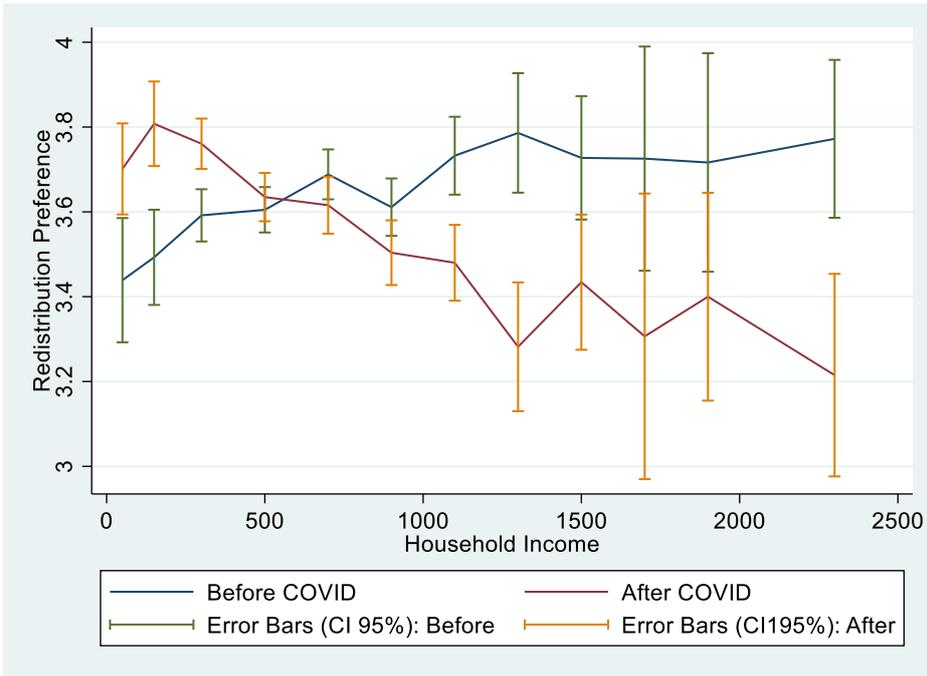

Fig. 5. Means of Redistribution Preference by household income before and after COVID 19 (Mean values of Residential Prefecture)

Note: A balance panel is used to illustrate the figures.

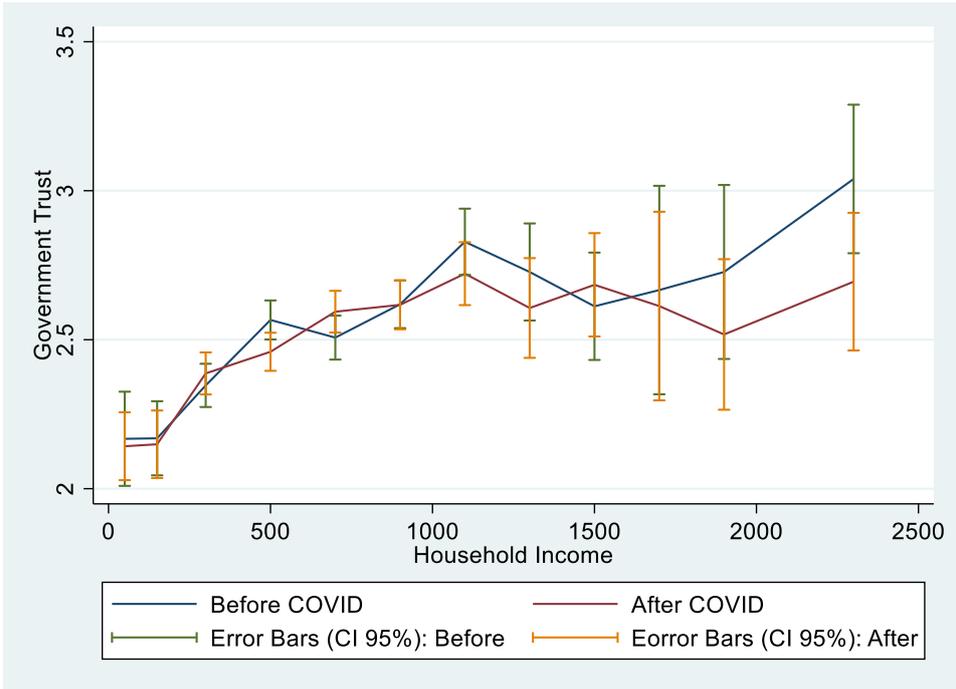

Fig. 6(a). Government trust

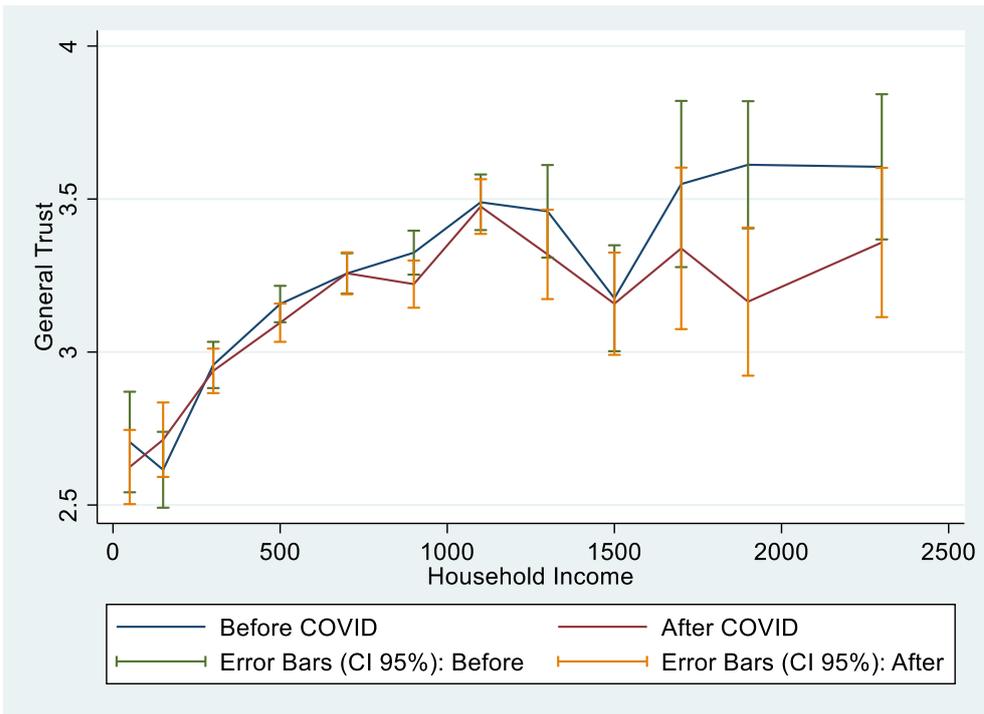

Fig. 6(b). Government trust

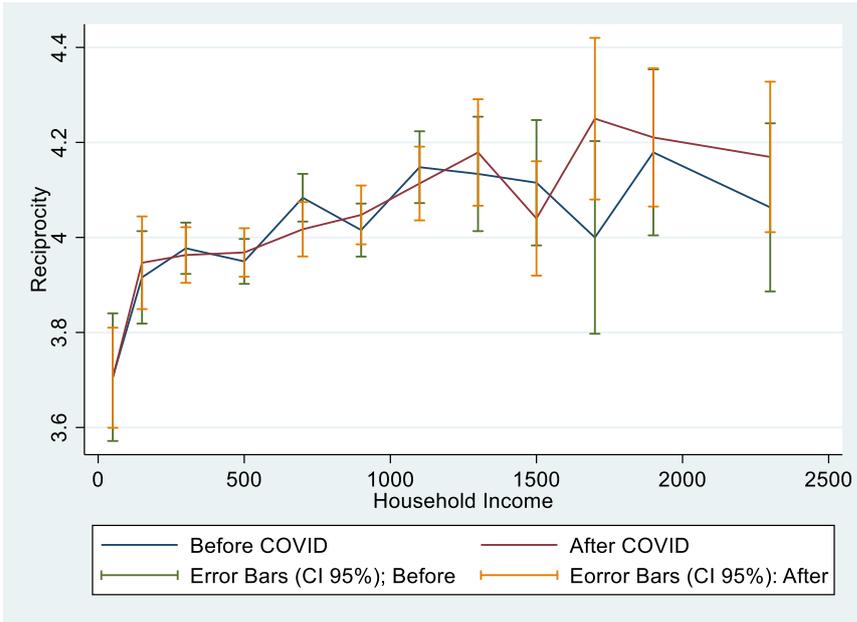

Fig. 6(c). Government trust

Fig. 6. Means of subjective values by household income before and after COVID-19 (Mean values of residential prefectures)

Note: A balance panel is used to illustrate the figures.

Table 1. Definitions of variables and mean values before and after COVID-19 (Balanced panel)

| Variables | Definition | Before COVID-19 | After COVID-19 |
|---|---|---|---|
| *Income* | Household income (million yen) | 0.36 | 0 |
| *High Income* | Equals 1 if respondent's household income is higher than its median values, otherwise 0. | 0.39 | 0.75 |
| *Covid* | Equals 1 for the survey conducted in 2021, 2023 or 2024, otherwise 0 for the survey in 2016, 2017, or 2018. | 0 | 1 |
| *Redistribution* | Question: 'The government should reduce the gap between the rich and poor' Respondents' values on a 5-point agreement scale (1 = strongly disagree, 5 = strongly agree). | | |
| *Government Trust_1* | Equals 1 if the respondent chose 1 (do not trust government at all), otherwise 0. | 0.20 | 0.28 |
| *Government Trust_2* | Equals 1 if the respondent chose 2 (do not trust government), otherwise 0. | | |
| *Government Trust_3* | Equals 1 if the respondent chose 3 (indecisive), otherwise 0. | | |
| *Government Trust_4* | Equals 1 if the respondent chose 4 (trust government), otherwise 0. | | |
| *Government Trust_5* | Equals 1 if the respondent chose 5 (trust government very much), otherwise 0. | 0.15 | 0.21 |
| *General Trust_1* | Equals 1 if the respondent chose 1 (do not trust others at all), otherwise 0. | 0.020 | 0.013 |
| *General Trust_2* | Equals 1 if the respondent chose 2 (do not trust others), otherwise 0. | 0.009 | 0.005 |
| *General Trust_3* | Equals 1 if the respondent chose 3 (indecisive), otherwise 0. | | |
| *General Trust_4* | Equals 1 if the respondent chose 4 (trust others), otherwise 0. | | |
| *General Trust_5* | Equals 1 if the respondent chose 5 (trust others very much), otherwise 0. | 0.39 | 0.38 |
| *Reciprocity_1* | Equals 1 if the respondent chose 1 (do not agree at all), otherwise 0. | 0.41 | 0.41 |
| *Reciprocity_2* | Equals 1 if the respondent chose 2 (do not agree), otherwise 0. | 0.44 | 0.42 |
| *Reciprocity_3* | Equals 1 if the respondent chose 3 (indecisive), otherwise 0. | 0.85 | 0.83 |
| *Reciprocity_4* | Equals 1 if the respondent chose 4 (agree), otherwise 0. | 0.59 | 0.49 |
| *Reciprocity_5* | Equals 1 if the respondent chose 5 (strongly agree), otherwise 0. | 44.7 | 44.1 |
| *Male* | Equals 1 if the respondent is male, otherwise 0. | | |
| *Age* | Respondent's male | | |
| *Married* | Equals 1 if the respondent is married, otherwise 0. | | |
| *University* | Equals 1 if the respondent's highest academic background is university graduate or above, otherwise 0. | | |

Note: In addition, we included 14 job status dummies and 46 residential prefecture dummies.

Table 2. Estimation results, FE model: Balanced panel

| | (1) | (2) | (3) | (4) | (5) |
|---|---|---|---|---|---|
| *Covid* ×*High Income* | −0.302*** (0.033) | −0.303*** (0.033) | −0.303*** (0.033) | −0.301*** (0.037) | −0.302*** (0.036) |
| *Covid* | 0.004 (0.038) | 0.0004 (0.038) | 0.005 (0.038) | −0.036 (0.041) | −0.038 (0.041) |
| *High Income* | 0.172*** (0.037) | 0.175*** (0.037) | 0.173*** (0.037) | 0.137*** (0.041) | 0.141*** (0.041) |
| *Government Trust_1* | <Default> | <Default> | <Default> | <Default> | <Default> |
| *Government Trust_2* | | 0.058* (0.032) | | | 0.083** (0.037) |
| *Government Trust_3* | | 0.057* (0.034) | | | 0.082** (0.030) |
| *Government Trust_4* | | 0.158*** (0.041) | | | 0.173** (0.048) |
| *Government Trust_5* | | 0.351*** (0.073) | | | 0.355*** (0.083) |
| *General Trust_1* | <Default> | | <Default> | | <Default> |
| *General Trust_2* | | | −0.111** (0.047) | | −0.149*** (0.052) |
| *General Trust_3* | | | −0.086* (0.050) | | −0.155*** (0.057) |
| *General Trust_4* | | | −0.034 (0.053) | | −0.132** (0.061) |
| *General Trust_5* | | | 0.053 (0.067) | | −0.087 (0.078) |
| *Reciprocity_1* | <Default> | | | <Default> | <Default> |
| *Reciprocity_2* | | | | 0.265** (0.129) | 0.275** (0.129) |
| *Reciprocity_3* | | | | 0.401*** (0.109) | 0.410*** (0.109) |
| *Reciprocity_4* | | | | 0.544*** (0.110) | 0.551*** (0.110) |
| *Reciprocity_5* | | | | 0.686*** (0.111) | 0.686*** (0.111) |
| Adj. R-square | 0.365 | 0.367 | 0.366 | 0.363 | 0.365 |
| Obs. | 10,390 | 10,374 | 10,375 | 8,690 | 8,672 |
| Individuals | 1,907 | 1,906 | 1,906 | 1,896 | 1,895 |

Note: In all columns, *Age, Male,* and *University* were controlled for because of the FE estimation. Additionally, 14 job status dummies and 46 residential prefecture dummies were included; however, these results were not reported. Numbers in parentheses are robust standard errors clustered by individuals. ***, **, and * indicate statistical significance at the 1%, 5%, and 10% levels, respectively.

Table 3. Estimation results for the FE model: Unbalanced Panel

|  | (1) | (2) | (3) | (4) | (5) |
|---|---|---|---|---|---|
| *Covid* ×*High Income* | −0.288*** (0.026) | −0.286*** (0.026) | −0.286*** (0.026) | −0.303*** (0.028) | −0.301*** (0.028) |
| *Covid* | −0.046 (0.028) | −0.052* (0.028) | −0.048* (0.028) | −0.074*** (0.029) | −0.080*** (0.029) |
| *High Income* | 0.084*** (0.025) | 0.084*** (0.025) | 0.083*** (0.025) | 0.070*** (0.027) | 0.070*** (0.027) |
| *Government Trust_1* | <Default> | <Default> |  |  | <Default> |
| *Government Trust_2* |  | 0.041* (0.021) |  |  | 0.061*** (0.023) |
| *Government Trust_3* |  | 0.045* (0.023) |  |  | 0.069*** (0.026) |
| *Government Trust_4* |  | 0.150*** (0.028) |  |  | 0.154*** (0.031) |
| *Government Trust_5* |  | 0.311*** (0.049) |  |  | 0.296*** (0.053) |
| *General Trust_1* | <Default> |  | <Default> |  | <Default> |
| *General Trust_2* |  |  | −0.072** (0.034) |  | −0.098*** (0.036) |
| *General Trust_3* |  |  | −0.056 (0.036) |  | −0.102*** (0.039) |
| *General Trust_4* |  |  | 0.004 (0.038) |  | −0.066 (0.042) |
| *General Trust_5* |  |  | 0.101** (0.047) |  | −0.001 (0.051) |
| *Reciprocity_1* | <Default> |  |  | <Default> | <Default> |
| *Reciprocity_2* |  |  |  | 0.156* (0.089) | 0.157* (0.089) |
| *Reciprocity_3* |  |  |  | 0.272*** (0.076) | 0.272*** (0.076) |
| *Reciprocity_4* |  |  |  | 0.395*** (0.076) | 0.391*** (0.076) |
| *Reciprocity_5* |  |  |  | 0.509*** (0.077) | 0.500*** (0.077) |
| Adj. R-square | 0.394 | 0.396 | 0.394 | 0.407 | 0.409 |
| Obs. | 27,084 | 27,036 | 27,046 | 24,966 | 24,912 |
| Individuals | 10,518 | 10,507 | 10,510 | 10,495 | 10,482 |

Note: In all columns, *Age, Male,* and *University* were controlled for through the Fixed Effects (FE) estimation. Additionally, 14 job status dummies and 46 residential prefecture dummies were included; however, their results were not reported. Numbers in parentheses are robust standard errors clustered by individuals. ***, **, and * indicate statistical significance at the 1%, 5%, and 10% levels, respectively.

Table 4. Estimation results, FE model: Balanced panel.

| | (1) | (2) | (3) | (4) |
|---|---|---|---|---|
| Covid ×High Income ×Government Trust_1 | <Default> | <Default> | <Default> | <Default> |
| Covid ×High Income ×Government Trust_2 | 0.080 (0.098) | 0.118 (0.104) | 0.103 (0.114) | 0.124 (0.121) |
| Covid ×High Income ×Government Trust_3 | 0.079 (0.091) | 0.119 (0.100) | 0.128 (0.105) | 0.152 (0.116) |
| Covid ×High Income ×Government Trust_4 | 0.186* (0.104) | 0.259** (0.117) | 0.199* (0.119) | 0.265** (0.134) |
| Covid ×High Income ×Government Trust_5 | 0.782*** (0.254) | 0.893*** (0.270) | 0.693** (0.290) | 0.834*** (0.312) |
| Covid ×High Income ×General Trust_1 | | <Default> | | <Default> |
| Covid ×High Income ×General Trust_2 | | −0.089 (0.145) | | 0.002 (0.167) |
| Covid ×High Income ×General t Trust_3 | | −0.079 (0.139) | | 0.011 (0.158) |
| Covid ×High Income ×General Trust_4 | | −0.164 (0.137) | | −0.076 (0.158) |
| Covid ×High Income ×General Trust_5 | | −0.240 (0.196) | | −0.213 (0.228) |
| Covid ×High Income ×Reciprocity_1 | | | <Default> | <Default> |
| Covid ×High Income ×Reciprocity_2 | | | −0.449 (0.520) | −0.464 (0.525) |
| Covid ×High Income ×Reciprocity_3 | | | −0.093 (0.432) | −0.082 (0.438) |
| Covid ×High Income ×Reciprocity_4 | | | −0.192 (0.427) | −0.165 (0.435) |
| Covid ×High Income ×Reciprocity_5 | | | −0.201 (0.428) | −0.167 (0.435) |
| *Linear combination* | | | | |
| Covid ×High Income+ Covid ×High Income ×Government Trust_4 | −0.17** (0.08) | −0.01 (0.15) | 0.01 (0.43) | 0.06 (0.44) |
| Covid ×High Income+ Covid ×High Income ×Government Trust_5 | 0.43* (0.24) | 0.62** (0.28) | 0.51 (0.51) | 0.63 (0.52) |
| Adj. R-square | 0.364 | 0.379 | 0.370 | 0.377 |
| Obs. | 10,374 | 10,371 | 8,675 | 8,672 |
| Individuals | 1,906 | 1,906 | 1,895 | 1,895 |

Note: In all columns, Age, Gender dummy, and Educational Background were controlled for through the Fixed Effects (FE) estimation. Additionally, 14 job status dummies and 46 residential prefecture dummies were included; however, these results were not reported. Numbers in parentheses are robust standard errors clustered by individuals. ***, **, and * indicate statistical significance at the 1%, 5%, and 10% levels, respectively.

Table 5. Estimation results for the FE model: unbalanced panel.

|  | (1) | (2) | (3) | (4) |
|---|---|---|---|---|
| *Covid ×High Income ×Government Trust_1* | <Default> | <Default> | <Default> | <Default> |
| *Covid ×High Income ×Government Trust_2* | 0.132* (0.074) | 0.105 (0.077) | 0.119 (0.085) | 0.085 (0.089) |
| *Covid ×High Income ×Government Trust_3* | 0.113 (0.070) | 0.084 (0.077) | 0.134 (0.080) | 0.101 (0.088) |
| *Covid ×High Income ×Government Trust_4* | 0.259*** (0.080) | 0.248*** (0.090) | 0.255*** (0.091) | 0.246** (0.102) |
| *Covid ×High Income ×Government Trust_5* | 0.685*** (0.199) | 0.644*** (0.213) | 0.620*** (0.227) | 0.609** (0.244) |
| *Covid ×High Income ×General Trust_1* |  | <Default> |  | <Default> |
| *Covid ×High Income ×General Trust_2* |  | 0.192* (0.114) |  | 0.251* (0.129) |
| *Covid ×High Income ×General t Trust_3* |  | 0.203* (0.109) |  | 0.257** (0.124) |
| *Covid ×High Income ×General Trust_4* |  | 0.121 (0.109) |  | 0.174 (0.124) |
| *Covid ×High Income ×General Trust_5* |  | 0.225 (0.151) |  | 0.223 (0.173) |
| *Covid ×High Income ×Reciprocity_1* |  |  | <Default> | <Default> |
| *Covid ×High Income ×Reciprocity_2* |  |  | 0.437 (0.387) | 0.304 (0.391) |
| *Covid ×High Income ×Reciprocity_3* |  |  | 0.649** (0.324) | 0.541 (0.328) |
| *Covid ×High Income ×Reciprocity_4* |  |  | 0.571* (0.320) | 0.469 (0.326) |
| *Covid ×High Income ×Reciprocity_5* |  |  | 0.608* (0.321) | 0.503 (0.326) |
| *Linear combination* |  |  |  |  |
| *Covid ×High Income+ Covid ×High Income ×Government Trust_4* | −0.11* (0.06) | −0.26** (0.12) | −0.72 (0.32) | −0.81 (0.33) |
| *Covid ×High Income+ Covid ×High Income ×Government Trust_5* | 0.31 (0.19) | 0.14 (0.22) | −0.36 (0.38) | −0.44 (0.39) |
| Adj. R-square | 0.403 | 0.404 | 0.416 | 0.417 |
| Obs. | 27,036 | 27,029 | 24,919 | 24,912 |
| Individuals | 10,507 | 10,505 | 10,484 | 10,482 |

Note: In all columns, Age, Gender dummy, and Educational Background were controlled for through the fixed effects (FE) estimation. Additionally, 14 job status dummies and 46 residential prefecture dummies were included; however, these results were not reported. Numbers in parentheses are robust standard errors clustered by individuals. ***, **, and * indicate statistical significance at the 1%, 5%, and 10% levels, respectively.